\documentclass[twocolumn,aps,floats,prl]{revtex4}
\usepackage{epsfig}
\begin{document}
\noindent {\bf Comment on "Mechanism of Terahertz Electromagnetic
Wave Emission from Intrinsic Josephson Junctions"}

In a recent Letter \cite{Tachiki09}, Tachiki \emph{et al.} argued
that the dynamic state of superconductivity phase in multi junctions
characterized by $\pm\pi$ kinks proposed in
Refs.~\cite{szlin08b,hu08,Koshelev08b,szlin09a} is hard to realize
from an energetic point of view. Instead, they proposed a so-called
"state without kink". In this Comment, we point out that their
energetic consideration is inconsistent with the well established
treatment for Josephson junctions, and that the results derived for
their "state without kink" cannot explain the radiations observed in
the recent experiment \cite{Ozyuzer07}.

Using an energetic analysis based on Ginzburg-Landau equation at equilibrium, the authors claimed that the $\pm\pi$
kinks will suppress the superconductivity amplitude, which has not been taken into account in
Refs.~\cite{szlin08b,hu08,Koshelev08b,szlin09a}. In order to clarify this point, we calculate the in-plane supercurrent
created by a $\pi$ kink. It is found that the maximal value of in-plane supercurrent caused by the kink is $\sim
100J_c$, with $J_c$ the critical current of the junction, which is too small compared to the in-plane critical current
density in order of $250000J_c$ (taking the anisotropy parameter $\gamma=500$) to suppress sensibly the
superconductivity amplitude. This estimate is consistent with the established treatment on soliton phenomena in
Josephson junctions; although the suppression of superconductivity amplitude has never been included in junction
physics, calculated \emph{IV} curves for soliton states agree well with experimental results. Therefore, the energetics
reported in Ref.~\cite{Tachiki09} is inappropriate for Josephson junctions.

We have also computed the static magnetic field associated with the kinks, and found that it is negligibly small
compared to the magnetic field associated with the Josephson plasma. Since the $\pm\pi$ kinks are alternatively aligned
along the stack direction, the energy cost of the total \textit{antiferromagnetic} structure is very small in
comparison to the kinetic energy and the energy carried by the plasma oscillation\cite{szlin09a}.

We also notice that energetics is not sufficient for understanding the dynamics of the present nonlinear system. It is
the stability that governs the non-equilibrium state. Fed by a dc energy input, the system does not have to stay in the
lowest energy state, instead it prefers a stable state. For a long single junction, the state with solitons of $2\pi$
phase kinks is stabilized over the McCumber state of uniform phase due to the parametric instability\cite{Pagano86} at
a sequence of bias voltages, even though a soliton carries higher energy. Similarly, for inductively coupled multi
junctions, the McCumber state evolves into the state with $\pm\pi$ phase kinks due to instability near cavity
resonances.

We then look into the so-called "state without kink", which Tachiki \emph{et al.} claimed responsible for the
experiment\cite{Tachiki09}. Unfortunately, the authors have not shown the phase configuration for this state (Fig.~2 is
for the $\pi$-kink state which they claimed unrealistic).

However, from the following observations, the state they obtained should be the soliton state. First, the EM fields in
Fig.~5(a) are similar to the ones associated with a soliton in a single junction \cite{szlin09a}. Secondly, in Fig.~4
the voltages where current steps occurred are precisely given by $V'=2n\pi/L_x'$ with $n$ being integer. Thirdly, the
slop in inverse-length dependence of voltage for the lowest current step shown in Fig.~6 is $\sim 2\pi$. All these
features coincide exactly with those of soliton solutions. Therefore, the current steps in Fig.~4 are caused by moving
solitons, and known for long time as zero-field steps.

It is well known that the ac Josephson relation never holds for the soliton state. Therefore, a soliton state cannot be
responsible for the radiation observed in the recent experiment where ac Josephson relation is
confirmed\cite{Ozyuzer07}. Solitons cannot be aligned uniformly along the stack direction due to the strong repulsion
force. The soliton state therefore conflicts with the experimental observation on a coherent radiation
\cite{Ozyuzer07}.

We also notice that all the results for their "state without kink"
are for $L'_x>1$ (see Figs.~4, 5 and 6), outside the regime explored
in experiment $L'_x=L_x/\lambda_c<0.5$ taking $\lambda_c=200\mu$m at
zero temperature\cite{Ozyuzer07}. If the absence of result means
their "state without kink" unstable for $L'_x<1$, it is again
compatible with the known property of soliton states, namely a
soliton cannot be stabilized when the junction length is smaller
than the penetration depth.

In contrast, the $\pi$ kink state discussed in Refs. \cite{szlin08b,hu08,Koshelev08b,szlin09a} is stable and supports
coherent THz radiations with frequency satisfying the ac Josephson relation.

This work was supported by WPI Initiative on Materials
Nanoarchitronics, MEXT, Japan, CREST-JST, Japan and partially by
ITSNEM of CAS.

\vspace{2mm}

\noindent
Shizeng Lin and Xiao Hu\\
\indent WPI Center for Materials Nanoarchitectonics,\\
\indent National Institute for Materials Science,\\
\indent Tsukuba 305-0044, Japan; \\
\indent Graduate School of Pure and Applied Sciences,\\
\indent University of Tsukuba, Tsukuba 305-8571, Japan;\\
\indent Japan Science and Technology Agency, \\
\indent 4-1-8 Honcho, Kawaguchi, Saitama 332-0012, Japan

\date{\today}
\maketitle
\end{document}